\documentclass[conference]{IEEEtran}
\IEEEoverridecommandlockouts
\usepackage{cite}
\usepackage{amsmath,amssymb,amsfonts}
\usepackage{algorithm}
\usepackage{algorithmic}
\usepackage{graphicx}
\usepackage{textcomp}
\usepackage{xcolor}
\usepackage[small]{caption}
\usepackage{booktabs}
\usepackage{multirow}
\usepackage{subfigure}
\def\BibTeX{{\rm B\kern-.05em{\sc i\kern-.025em b}\kern-.08em
    T\kern-.1667em\lower.7ex\hbox{E}\kern-.125emX}}
\begin{document}

\title{Area Queries Based on Voronoi Diagrams\\
\thanks{The work is supported by the National Natural Science Foundation of China (U1711267, 41572314, 41972306), Hubei Province Innovation Group Project (2019CFA023) and Fundamental Research Funds for the Central Universities, China University of Geosciences (Wuhan) (CUGCJ1810).}
}

\author{\IEEEauthorblockN{Yang Li}
\IEEEauthorblockA{
Supervised by Gang Liu\\
\textit{School of Computer Science} \\
\textit{China University of Geosciences}\\
Wuhan, China \\
liyang\_cs@cug.edu.cn}
}

\maketitle

\begin{abstract}
Area query, to find all elements contained in a specified area from  a  certain set of spatial objects, is a very important spatial query widely required in various fields.
A number of approaches have been proposed to implement this query, the best known of which is to obtain a rough candidate set through spatial indexes and then refine the candidates through geometric validations to get the final result.
When the shape of the query area is a rectangle, this method has very high efficiency. However, when the query area is irregular, the candidate set is usually much larger than the final result set, which means a lot of redundant detection needs to be done, thus the efficiency is greatly limited.
In view of this issue, we propose a method of iteratively generating candidates based on Voronoi diagrams and apply it to area queries.
The experimental results indicate that with our approach, the number of candidates in the process of area query is greatly reduced and the efficiency of the query is significantly improved.
\end{abstract}


\section{Introduction}
Spatial query\cite{Orenstein:1986:SQP:16894.16886,DBLP:conf/gis/2016,Implementing_Spatial_Data_Analysis_Software_Tools_in_R,Bivand2000}, to retrieve the corresponding spatial object set from the spatial database according to the given spatial constraints, is one of the most commonly used functions in geographic information system (GIS). 
As the basis of spatial analysis, it is widely used in various fields, such as aerospace, military, physics, urban planning, transportation planning, logistics management, etc\cite{DBLP:journals/access/MingLPWG18,Scott2010,DBLP:conf/ijcnn/LiLLL17,geostar,ogc}.
For GIS with large-scale data, the efficiency of  its spatial query directly affects the performance of the entire application. 
The area query is a typical and very important type of spatial queries that requires to find all spatial objects in a given closed spatial area from the database.

\begin{figure*}[htbp]
 \centering
 \subfigure[Traditional method]{\includegraphics[width=0.49\textwidth]{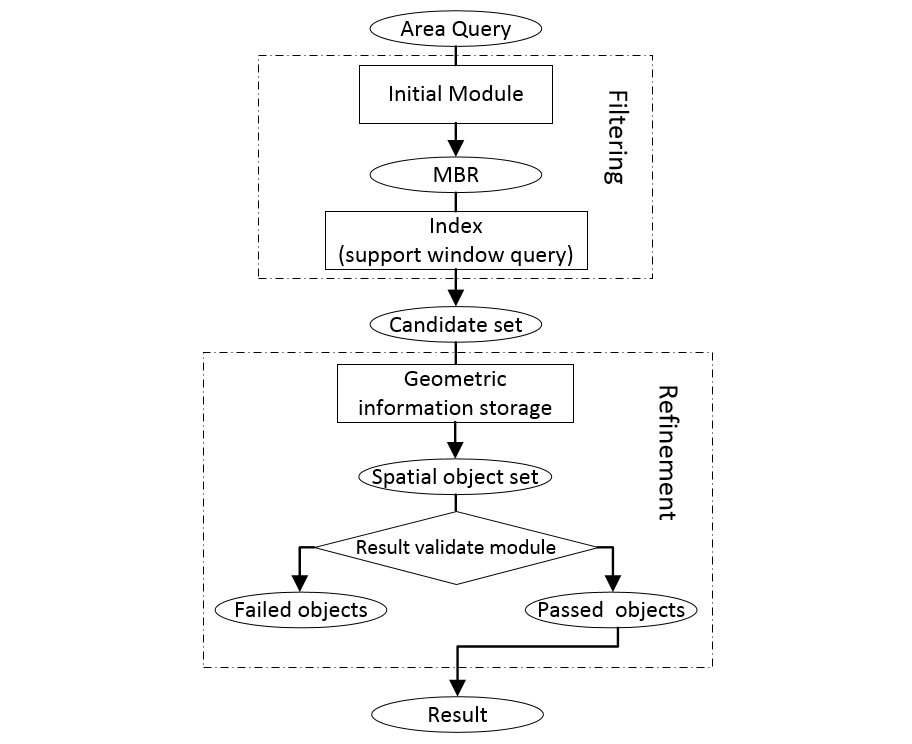}}
 \subfigure[Our method]{\includegraphics[width=0.49\textwidth]{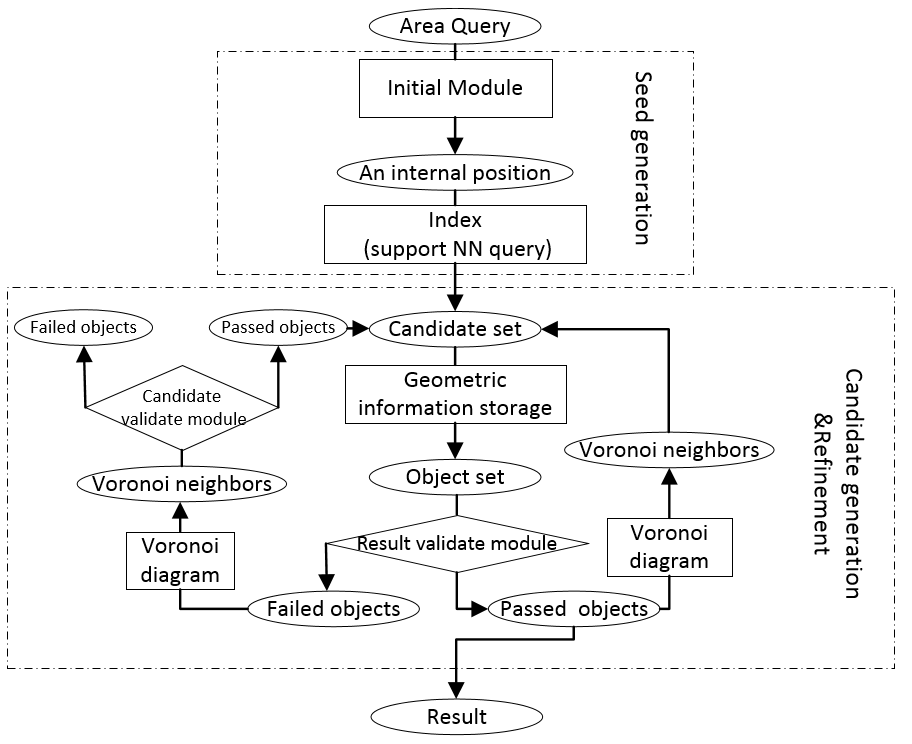}}
 \caption{The processes of traditional approach and our approach for the area query}
 \label{pro_compare}
\end{figure*}

Like other spatial queries, the area query is not only computationally intensive, but also IO intensive.
Therefore,  an effective way to improve the efficiency of area query is to reduce both IO complexity either computational complexity in the query process.
For this purpose, a great number of approaches have been proposed, the best known of which is to obtain a rough candidate set through spatial indexes and then refine the candidate set through geometric verification modules to get the final result.
 As shown in Fig.\,\ref{pro_compare} (a), the process of this traditional implementation of the area query consists of two steps, i.e., filtering and refining.
The first step in this process does not require loading the full geometry information or performing expensive geometry calculations, just efficient index retrieval. However, the vast majority of objects that do not meet the query criteria can be filtered out.
The second step only requires further refinement of the results from the previous step. It is usually more time consuming than the first step because of its geometric information loading and complex geometric calculations.
For a long time, a large number of researchers have been trying to design better indexing structures to improve the retrieval efficiency of the filtering process or propose a more efficient geometric verification algorithm to reduce the time cost in the process of refinement.
Many spatial indexes and their improvements have been proposed, among which the most classical ones are R-tree\cite{Guttman:1984:RDI:602259.602266,Performance_Evaluation_of_Main-Memory_R_tree_Variants,Sharifzadeh:2010:VRV:1920841.1920994,Cheung:1998:ENN:290593.290596,Ang:1997:NLN:647225.718938,Arge:2004:PRP:1007568.1007608,Manolopoulos:2005:RTA:1098699}, quadtree\cite{Samet1984The,Shekhar2008,8151505}, KD-tree\cite{Bentley:1975:MBS:361002.361007,Friedman:1977:AFB:355744.355745,Procopiuc2003Bkd,DBLP:conf/sigmod/Robinson81}, etc.
Many companies and open source organizations such as OGC\cite{ogc}, ESRI\cite{Scott2010} and GeoStar\cite{geostar} have launched their spatial data engines, which can usually provide very efficient spatial geometry computation.
For the traditional area query approach, mentioned above, the spatial index must support window query (also known as range query), and the candidate set generated by the filtering process is obtained by window query on the spatial index through the minimum bounding rectangle (MBR) of the query area. So the candidate set contains all the objects in the MBR of the query area.
If the shape of the query area is very similar to a rectangle, the area of the query area will be very close to that of its MBR, the result set will also be very close to the candidate set in size.
However, if the shape of the query area is irregular, the area of the query area is often much smaller than that of its MBR. For example, the area of a triangle must not be greater than half of the area of its MBR.
In this case, the size of the candidate set in the query process will be much larger than the size of the result set.
In practice, the shape of the query area given by users is often an irregular polygon, more often a concave polygon.
As a result, the candidate set obtained by a window query is usually much larger compared with the final result set.

\begin{figure}[htbp]
 \centering
 \subfigure[Traditional method]{\includegraphics[width=0.24\textwidth]{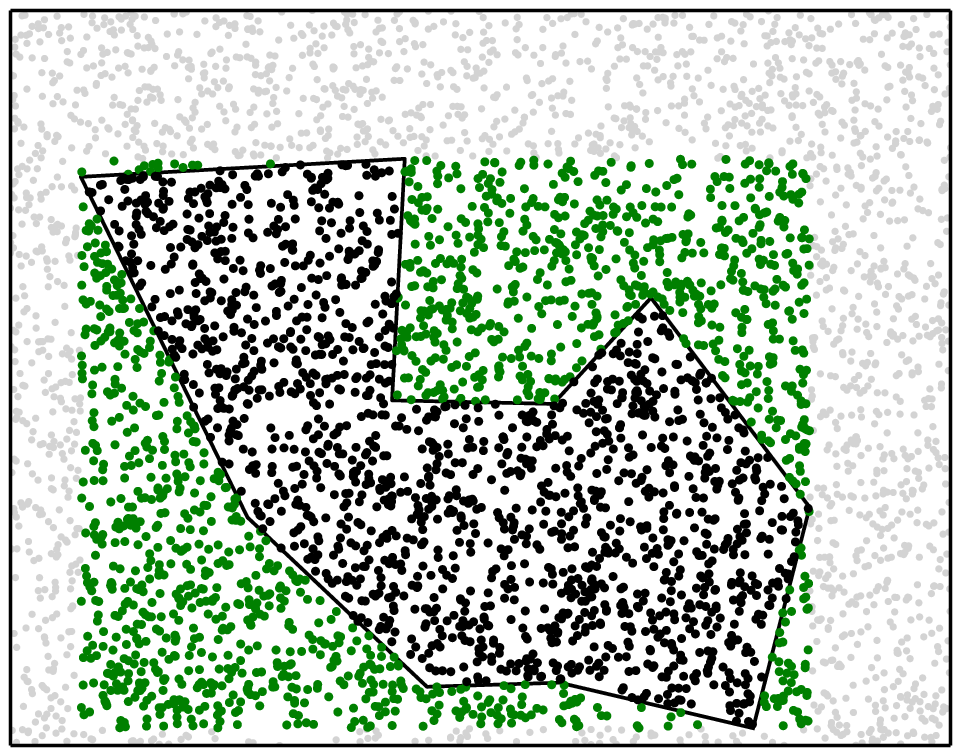}}
 \subfigure[Our method]{\includegraphics[width=0.24\textwidth]{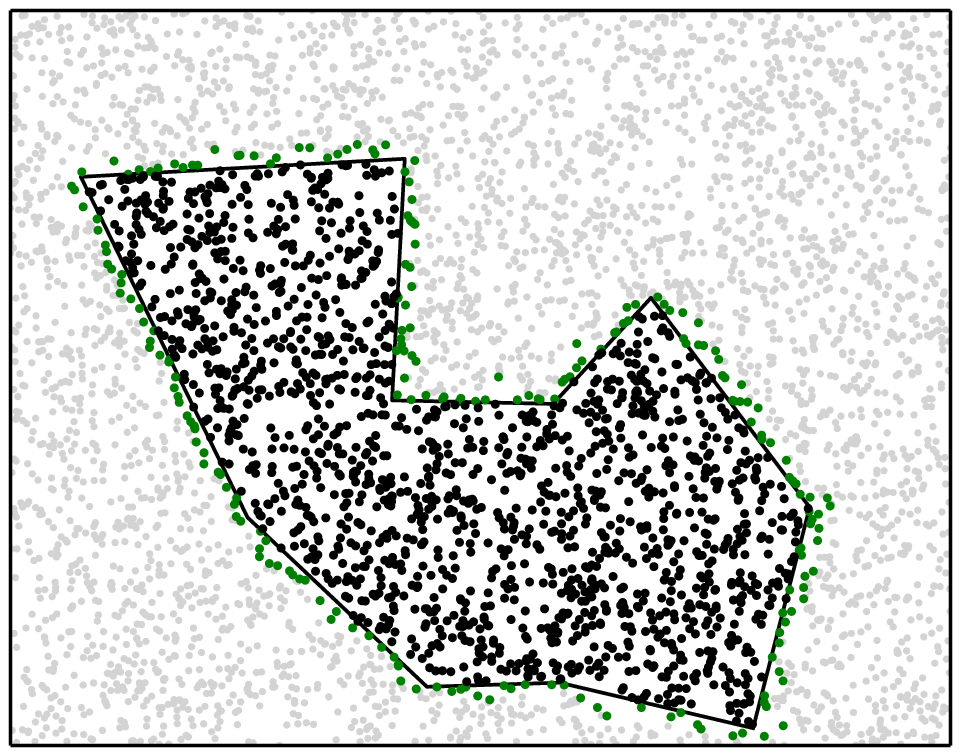}}
 \caption{The points in the result set and candidate set of the area query, where the query area is enclosed by black lines, black points belong to the result set and green points belong to the candidate set.}
 \label{point_compare}
\end{figure}

If the candidates can be reduced, the time cost of the refining process will be reduced, so that the efficiency of the entire area query will be improved.
Obviously, for an area query, there are many adjacency relationships among all objects of the result set. In other words, if an object is in a given area, its adjacent objects are likely to be in the same area.
As a kind of spatial partition, the Voronoi diagram can clearly show the adjacency relationships among spatial objects.
Based on this idea, We propose a novel area query implement approach as shown in Fig.\,\ref{pro_compare}\,(b), in which the candidate set is not directly obtained through a window query, but generated by an incremental algorithm based on Voronoi diagrams.
The candidate set generated by this incremental algorithm only contains all points in the query area and a small number of points out of the query area but close to the boundary of the query area. It is generally much smaller than the candidate set obtained by window queries, as shown in Fig.\,\ref{point_compare}. Hence, the proposed area query method based on the Voronoi diagram performs more efficiently than the traditional one.

\section{Background}
\begin{figure}[htbp]
    \centering
    \includegraphics[width=0.49\textwidth]{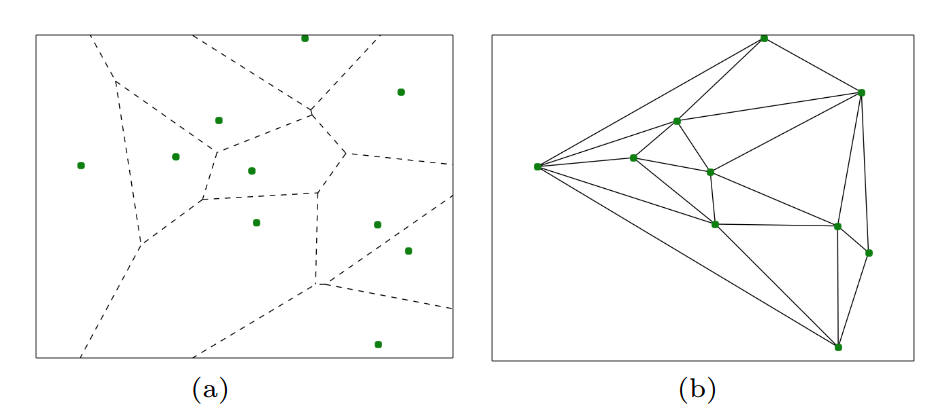}
    \caption{a) Voronoi diagram, b) Delaunay triangulation}
    \label{vd_dt}
\end{figure}

The background on the Voronoi diagram and  Delaunay triangulation is introduced in order to have a better understanding of the mechanism of our approach. A Voronoi diagram and a Delaunay triangulation is illustrated as Fig.\,\ref{vd_dt}.
\subsection{Voronoi diagram and its properties}

The Voronoi diagram\cite{DBLP:books/wi/OkabeBSCK00,DBLP:reference/db/ShahabiS18,DBLP:reference/gis/ShahabiS17}, proposed by Rene Descartes in 1644, is a spatial partition structure widely applied in many science domains, most notably spatial database and computational geometry. In a Voronoi diagram of $n$ points, the space is divided into $n$ regions corresponding to these points, which are called Voronoi cells.
For each point of these $n$ points, the corresponding Voronoi cell consists of all locations closer to that point than to any other.
In other words, each point is the closest point to all the locations in its corresponding Voronoi cell.
Formally, the above description can be stated as follows.
Given a set $P$ of $n$ points, the Voronoi cell of a point $p \in P$, written as $V(P, p)$, is defined as Eq. \eqref{eq4}, where $\parallel p-q\parallel$ represents the Euclidean distance between $p$ and $q$.
\begin{align}
V(P, p)=\{q\;|\;
\forall p' \in P,\; p' \neq p,\;\parallel p-q\parallel \leq \parallel p'-q\parallel\}
    \label{eq4}
\end{align}
The Voronoi diagram of $P$, written as $VD(P)$, is defined as Eq. \eqref{eq5}.
\begin{align}
VD(P)=\{V(P, p)\;|\; p \in P\}
    \label{eq5}
\end{align}
The Voronoi neighbors of a point $p \in P$ is written as $VN(P, p)$.
The Voronoi diagram has the following properties:\\
\textbf{Property\,1}: The Voronoi diagram of a certain set $P$ of points, $VD(P$), is unique.\\
\textbf{Property\,2}: Given the Voronoi diagram of $P$, the nearest point of $P$ to a point $q\in P$ is among the Voronoi neighbors of $q$. That is, the closest point to $q$ is one of the generator points whose Voronoi cells share a Voronoi edge with $V(P,q)$. \\
\textbf{Property\,3}: Given the Voronoi diagram of $P$ and a test point $q \notin P$, a point $p'$ is the nearest point of $P$ to $p$, if and only if $q \in V(P,p')$.\\

\subsection{Delaunay triangulation and its properties}

Delaunay triangulation\cite{Delaunay1934Sur} is a very famous triangulation proposed by Boris  Delaunay in 1934. For a set $P$  of discrete points in a plane, the Delaunay triangulation $DT(P)$ is such a triangulation that no point in $P$ is inside the circumcircle of any triangle of $DT(P)$. 
The Delaunay  triangulation has the following properties:
\\
\textbf{Property\,4}: The Delaunay triangulation of a set of points is dual to its Voronoi diagram.\\
\textbf{Property\,5}: A graph of Delaunay triangulation must be a connected graph, that is, any two vertices in the graph are connected.\\
\textbf{Property\,6}: For a set of points, its nearest neighbor graph is a subgraph of its Delaunay triangulation graph.\\

\section{Methodologies}
In this section, the implementation and mechanism of our proposed Voronoi diagram based area query are described in detail.

Given a query area $A$ and a set $P$ of points, the points in $P$ can be divided into three types:
\begin{itemize}
    \item \textbf{Internal points}: the points in $P$ contained in $A$.
    \item \textbf{External points}: the points in $P$ not contained in $A$ and not adjacent to the boundary of $A$.
    \item \textbf{Boundary points}: the points in $P$ not contained in $A$ but adjacent to the boundary of $A$.
\end{itemize}

From the properties of Voronoi diagram and Delaunay graph, the following properties about the three types of points are obtained.\\
\textbf{Property\,7}: For any internal point of $A$, each of its Voronoi neighbors must be an internal or boundary point of $A$.\\
\textbf{Property\,8}: For any external point of $A$, each of its Voronoi neighbors must be an external or boundary point of $A$.\\
\textbf{Property\,9}: For any boundary point $p$ of $A$, there must exist at least one such point $q$, that the line between $p$ and $q$ intersects $A$.

\begin{algorithm}[ht]
\caption{Voronoi Diagram Based Area Query}
\label{pseudocode}
\textbf{Input}: The point set $P$ and the query area $A$\\
\textbf{Output}: The set $P_{result}$ of all the points in $A$ of $P$

\begin{algorithmic}[1]
\STATE $P_{candidate} :=$ an empty queue;
\STATE $P_{result} :=$ an empty queue;
\STATE $p_A :=$ an arbitrary position in $A$;
\STATE $p_{seed} :=$ NN($P$, $p_A$);
\STATE Append($P_{candidate}$, $p_{seed}$)
\STATE $P_{visited} := \{p_{seed}\}$
\WHILE{Size($P_{candidate}$) $> 0$}
\STATE $p := $Pop($P_{candidate}$);
\IF{Contains($A$, $p$)}
\STATE Append($P_{result}$, $p$)
\FORALL{$p_n \in$ VN($P$, $p$)}
\IF{$p_n \notin P_{visited}$}
\STATE Append($P_{candidate}$, $p_n$)
\STATE $P_{visited} := P_{visited} \cup \{p_n\}$;
\ENDIF
\ENDFOR
\ELSE
\FORALL{$p_n \in$ VN($p$)}
\IF{$p_n \notin P_{visited}$}  
\STATE $line_{p,p_n} :=$ Line($p$, $p_n$);
\IF{Intersects($line_{p,p_n}$, $A$)}
\STATE Append($P_{candidate}$, $p_n$)
\STATE $P_{visited} := P_{visited} \cup \{p_n\}$;
\ENDIF
\ENDIF
\ENDFOR
\ENDIF
\ENDWHILE
\STATE \textbf{return} $P_{result}$.
\end{algorithmic}
\end{algorithm}
According to the above three properties, we can further conclude that any internal point cannot be adjacent to an external point.
In the Delaunay graph of $P$, there must be at least one connection path between any two internal or boundary points that only goes through the internal or boundary points.
Therefore, with the Voroni diagram or the Delaunay graph of $P$, we can start at any internal or boundary point of $A$ and access all the internal and boundary points one by one, without passing through any external points.
In general, the total number of internal and boundary points of $A$ is much smaller than the number of points contained in the MBR of $A$.
If all the internal and boundary points of the query area are used as the candidate set, a large number of IOs and geometric calculations can be reduced.
Based on this idea, we propose a Voronoi Diagram Based Area Query algorithm whose pseudo-code is shown in Algorithm \ref{pseudocode}.

First, find the nearest neighbor (NN) of an arbitrary position in the query area from the database through the spatial index, and put it into the candidate set as a seed point. With the properties of the Voronoi diagram, we can know that this seed point must either be an internal point or a boundary point of the query area.
Then take a point from the candidate set and determine its type. If it is an internal point of the query area, place it in the result set and all its unaccessed Voronoi neighbors in the candidate set. If it is a boundary point, only place its unaccessed Voronoi neighbors in the candidate set. Repeat this step until the candidate set is empty. Finally, the result set is returned.

\section{Experiments}
 In order to investigate the effectiveness of our method, we conduct a number of experiments with the traditional area query method based on R-tree as a comparison method.
 All the programs in the experiment are implemented by Python and run on a personal computer with the Python version of 2.7, the CPU as Intel Core i5-4308U 2.80GHz and the RAM as DDR3 8G.
 For fairness, the index used to provide the NN query in our method is also R-tree.
 To decrease the error of the experiments, we repeat each experiment for 1000 times and calculate the average of the results. The query area for each time of the experiment is a randomly generated polygon of ten points.
 
\begin{figure}[htbp]
    \centering
    \includegraphics[width=0.45\textwidth]{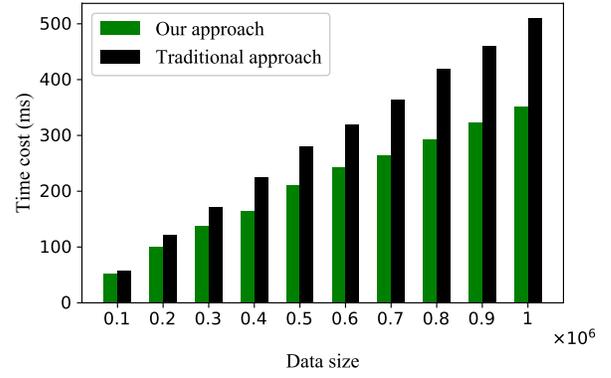}
    \caption{Time cost of area queries from data with various sizes}
    \label{data_size1}
\end{figure}

\begin{figure}[htbp]
    \centering
    \includegraphics[width=0.45\textwidth]{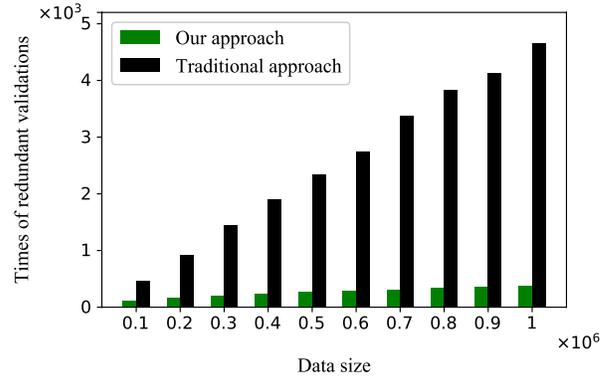}
    \caption{Times of redundant validations of area queries from data with various sizes}
    \label{data_size2}
\end{figure}

\begin{figure}[htbp]
    \centering
    \includegraphics[width=0.45\textwidth]{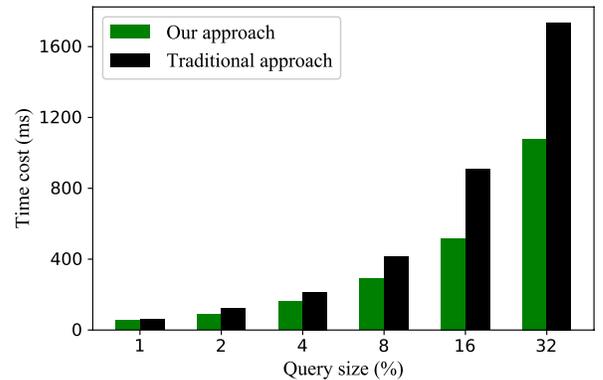}
    \caption{Time cost of area queries with various query sizes}
    \label{query_size1}
\end{figure}

\begin{figure}[htbp]
    \centering
    \includegraphics[width=0.45\textwidth]{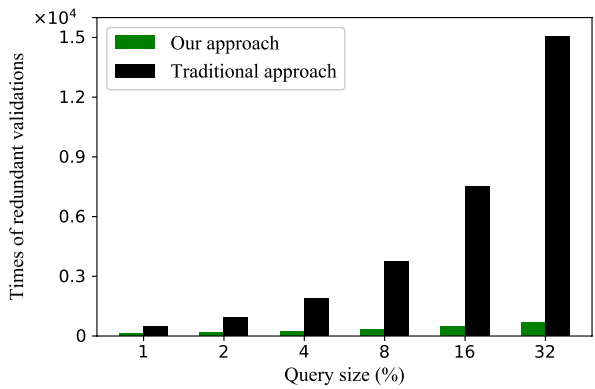}
    \caption{Times of redundant validations of area queries with various query sizes}
    \label{query_size2}
\end{figure}

Our experiments are designed into two sets.
The first set of experiments is used to evaluate the effect of data size on the performance of the area query algorithm.
The data size, number of points contained in the database, is from 1E5 to 1E6. The query size, i.e., the area of the query area's MBR divided by the total area of the solution space, is fixed at 1\%. 
Fig.\,\ref{data_size1} and  Fig.\,\ref{data_size2} show the time cost and the  validation times of redundant candidate at various data sizes, respectively. 
As the data size changes from 1E5 to 1E6, the time cost and the number of candidates saved by our method compared to the traditional one varies respectively from 10.6\% to 31.3\% and from 35.1\% to 42.9\%. 
The detailed experimental results are presented in Table \ref{T1}.
 \begin{table}[htbp]
\centering
\caption{The result size and the candidate set size and query time of R-tree based method and Voronoi diagrams based method with various sizes of data set}
\label{T1}
\resizebox{0.49\textwidth}{!}{
\begin{tabular}{@{}rrrrrr@{}}
\toprule
\multirow{2}{*}{\textbf{Data size}} & \multirow{2}{*}{\textbf{Result size}} & \multicolumn{2}{c}{\textbf{Traditional approach}}       & \multicolumn{2}{c}{\textbf{Our approach}}           \\ \cmidrule(lr){3-4} \cmidrule(l){5-6}
                                    &                                       & \textbf{Candidate number} & \textbf{Time (ms)} & \textbf{Candidate number} & \textbf{Time (ms)} \\ \midrule
1E5                                 & 529.34                                & 999.2                     & 58.389        & 648.47                    & 52.223        \\
2E5                                 & 1071.84                               & 1992.87                   & 122.077       & 1240.76                   & 100.022       \\
3E5                                 & 1557.27                               & 2998.52                   & 172.546       & 1763.57                   & 138.634       \\
4E5                                 & 2100.29                               & 3998.85                   & 225.714       & 2337.58                   & 163.775       \\
5E5                                 & 2645.02                               & 4990.78                   & 280.255       & 2912.21                   & 211.149       \\
6E5                                 & 3245.19                               & 5995.76                   & 320.078       & 3537.68                   & 242.718       \\
7E5                                 & 3615.22                               & 6997.33                   & 364.631       & 3930.04                   & 264.556       \\
8E5                                 & 4156.46                               & 7993.18                   & 419.281       & 4494.21                   & 292.487       \\
9E5                                 & 4861.76                               & 8998.72                   & 460.117       & 5220.23                   & 323.592       \\
1E6                                 & 5335.23                               & 9994.04                   & 511.066       & 5710.64                   & 351.086       \\ \bottomrule
\end{tabular}
}
\end{table}
The second set of experiments is used to evaluate effect of query size on the performance of the area query algorithm. The data size of this set is fixed at 1E5, and the query size varies form 1\% to 32\%.
Fig.\,\ref{query_size1} and  Fig.\,\ref{query_size2} show the time cost and the  validation times of redundant candidate at various query sizes respectively. 
As the query size changes from 1\% to 32\%, the time cost and the number of candidates saved by our method compared to the traditional one varies respectively from 11.7\% to 37.9\% and from 35.1\% to 44.9\%. 
The detailed experimental results are presented in Table \ref{T2}. 
\begin{table}[htbp]
\centering
\caption{The result size and the candidate set size and query time of R-tree based method and Voronoi diagrams based method with various sizes of queries}
\label{T2}
\resizebox{0.49\textwidth}{!}{ 
\begin{tabular}{@{}rrrrrr@{}}
\toprule
\multirow{2}{*}{\textbf{Query size}} & \multirow{2}{*}{\textbf{Result size}} & \multicolumn{2}{c}{\textbf{Traditional approach}}       & \multicolumn{2}{c}{\textbf{Our approach}}           \\ \cmidrule(lr){3-4} \cmidrule(l){5-6}
                                    &                                       & \textbf{Candidate number} & \textbf{Time (ms)} & \textbf{Candidate number} & \textbf{Time (ms)} \\ \midrule
1\%                                 & 529.34                                & 999.2                     & 60.44         & 648.47                    & 53.34         \\
2\%                                 & 1055.87                               & 1996.22                   & 120.634       & 1222.76                   & 88.705        \\
4\%                                 & 2115.12                               & 3985.7                    & 213.636       & 2350.06                   & 162.051       \\
8\%                                 & 4228.77                               & 7975.37                   & 413.57        & 4562.14                   & 290.46        \\
16\%                                & 8452.97                               & 15961.04                  & 908.995       & 8924.16                   & 517.05        \\
32\%                                & 16916.88                              & 31946.53                  & 1737.119      & 17585.94                  & 1078.718      \\ \bottomrule
\end{tabular}
}
\end{table}
 
 It can be seen from the above results that our area query method is effective in avoiding redundant validation candidates and reducing the time cost. Compared with the traditional method, our method has a more significant advantage for large data size and large query area. This is because the number of redundant candidates verified by our method is proportional to the boundary length of the query area, whereas the number of redundant candidates verified by the traditional method is determined by the area difference between the MBR of the query area and the query area itself. In general, there are far fewer redundant candidates verified by our methods than by the traditional method. The more points in the query area, the more significant our method advantage will be.

\section{Conclusions}

Based on our review, we find the weaknesses of the traditional implementation of area queries.
That is, faced with an irregular query area, especially when the area of the query area is much smaller than that of its MBR, the efficiency of the traditional area query method is quite limited.
In view of this issue, we propose a method of iteratively generating candidate set based on Voronoi diagrams and apply it to area queries.
Because the candidate set is greatly reduced, the proposed Voronoi diagrams based area query approach performs less spatial calculation than the traditional one. Therefore, the query efficiency of our approach is significantly higher in the vast majority of cases (except where the query area is very close to a rectangle). Based on the R-tree index, our method and the traditional method are implemented in Python. The comparative experiments are conducted between these two methods, and the experimental results show that our method is more efficient, especially when the size of the database and the size of the query area are large.
\bibliographystyle{unsrt}
\bibliography{paper}

\begin{thebibliography}{10}

\bibitem{Orenstein:1986:SQP:16894.16886}
Jack~A. Orenstein.
\newblock Spatial query processing in an object-oriented database system.
\newblock In {\em Proceedings of the 1986 ACM SIGMOD International Conference
  on Management of Data}, SIGMOD '86, pages 326--336, New York, NY, USA, 1986.
  ACM.

\bibitem{DBLP:conf/gis/2016}
Siva Ravada, Mohammed~Eunus Ali, Shawn~D. Newsam, Matthias Renz, and Goce
  Trajcevski, editors.
\newblock {\em Proceedings of the 24th {ACM} {SIGSPATIAL} International
  Conference on Advances in Geographic Information Systems, {GIS} 2016,
  Burlingame, California, USA, October 31 - November 3, 2016}. {ACM}, 2016.

\bibitem{Implementing_Spatial_Data_Analysis_Software_Tools_in_R}
Roger Bivand.
\newblock Implementing spatial data analysis software tools in r.
\newblock {\em Geographical Analysis}, 38(1):23--40, 2006.

\bibitem{Bivand2000}
Roger Bivand and Albrecht Gebhardt.
\newblock Implementing functions for spatial statistical analysis using the
  language.
\newblock {\em Journal of Geographical Systems}, 2(3):307--317, Sep 2000.

\bibitem{DBLP:journals/access/MingLPWG18}
Zi~Ming, Yang Li, Shijie Peng, Xuechao Wu, and Jinyi Guo.
\newblock Selection based on colony fitness for differential evolution.
\newblock {\em {IEEE} Access}, 6:78333--78341, 2018.

\bibitem{Scott2010}
Lauren~M. Scott and Mark~V. Janikas.
\newblock {\em Spatial Statistics in ArcGIS}, pages 27--41.
\newblock Springer Berlin Heidelberg, Berlin, Heidelberg, 2010.

\bibitem{DBLP:conf/ijcnn/LiLLL17}
Yang Li, Chengjun Li, Gang Liu, and Wei Long.
\newblock Fitness with diversity information for selection of evolutionary
  algorithms.
\newblock In {\em 2017 International Joint Conference on Neural Networks,
  {IJCNN} 2017, Anchorage, AK, USA, May 14-19, 2017}, pages 191--197, 2017.

\bibitem{geostar}
B.~{Lambrigtsen}, T.~{Gaier}, A.~{Tanner}, P.~{Kangaslahti}, and S.~{Brown}.
\newblock {GeoSTAR}.
\newblock In {\em AGU Fall Meeting Abstracts}, volume 2006, pages IN21A--1203,
  Dec 2006.

\bibitem{ogc}
A.~Rao, G.~S. Percivall, and Yonsook Enloe.
\newblock Overview of the ogc catalog interface specification.
\newblock In {\em Geoscience and Remote Sensing Symposium, 2000. Proceedings.
  IGARSS 2000. IEEE 2000 International}, 2000.

\bibitem{Guttman:1984:RDI:602259.602266}
Antonin Guttman.
\newblock R-trees: A dynamic index structure for spatial searching.
\newblock In {\em Proceedings of the 1984 ACM SIGMOD International Conference
  on Management of Data}, SIGMOD '84, pages 47--57, New York, NY, USA, 1984.
  ACM.

\bibitem{Performance_Evaluation_of_Main-Memory_R_tree_Variants}
Sangyong Hwang, Keunjoo Kwon, Sang~K. Cha, and Byung~S. Lee.
\newblock Performance evaluation of main-memory r-tree variants.
\newblock In Thanasis Hadzilacos, Yannis Manolopoulos, John Roddick, and Yannis
  Theodoridis, editors, {\em Advances in Spatial and Temporal Databases}, pages
  10--27, Berlin, Heidelberg, 2003. Springer Berlin Heidelberg.

\bibitem{Sharifzadeh:2010:VRV:1920841.1920994}
Mehdi Sharifzadeh and Cyrus Shahabi.
\newblock Vor-tree: R-trees with voronoi diagrams for efficient processing of
  spatial nearest neighbor queries.
\newblock {\em Proc. VLDB Endow.}, 3(1-2):1231--1242, September 2010.

\bibitem{Cheung:1998:ENN:290593.290596}
King~Lum Cheung and Ada Wai-Chee Fu.
\newblock Enhanced nearest neighbour search on the r-tree.
\newblock {\em SIGMOD Recod}, 27(3):16--21, September 1998.

\bibitem{Ang:1997:NLN:647225.718938}
Chuan-Heng Ang and T.~C. Tan.
\newblock New linear node splitting algorithm for r-trees.
\newblock In {\em Proceedings of the 5th International Symposium on Advances in
  Spatial Databases}, SSD '97, pages 339--349, London, UK, UK, 1997.
  Springer-Verlag.

\bibitem{Arge:2004:PRP:1007568.1007608}
Lars Arge, Mark de~Berg, Herman~J. Haverkort, and Ke~Yi.
\newblock The priority r-tree: A practically efficient and worst-case optimal
  r-tree.
\newblock In {\em Proceedings of the 2004 ACM SIGMOD International Conference
  on Management of Data}, SIGMOD '04, pages 347--358, New York, NY, USA, 2004.
  ACM.

\bibitem{Manolopoulos:2005:RTA:1098699}
Yannis Manolopoulos, Alexandros Nanopoulos, Apostolos~N. Papadopoulos, and
  Yannis Theodoridis.
\newblock {\em R-Trees: Theory and Applications}.
\newblock Springer Publishing Company, Incorporated, 2005.

\bibitem{Samet1984The}
Hanan Samet.
\newblock The quadtree and related heiarchical data structures.
\newblock {\em ACM Computing Surveys}, 16:187--260, 1984.

\bibitem{Shekhar2008}
Shashi Shekhar and Hui Xiong.
\newblock {\em Quadtree}, pages 931--931.
\newblock Springer US, Boston, MA, 2008.

\bibitem{8151505}
J.~{Tayeb}, Ö. {Ulusoy}, and O.~{Wolfson}.
\newblock A quadtree-based dynamic attribute indexing method.
\newblock {\em The Computer Journal}, 41(3):185--200, Jan 1998.

\bibitem{Bentley:1975:MBS:361002.361007}
Jon~Louis Bentley.
\newblock Multidimensional binary search trees used for associative searching.
\newblock {\em Commun. ACM}, 18(9):509--517, September 1975.

\bibitem{Friedman:1977:AFB:355744.355745}
Jerome~H. Friedman, Jon~Louis Bentley, and Raphael~Ari Finkel.
\newblock An algorithm for finding best matches in logarithmic expected time.
\newblock {\em ACM Trans. Math. Softw.}, 3(3):209--226, September 1977.

\bibitem{Procopiuc2003Bkd}
Octavian Procopiuc, Pankaj~K. Agarwal, Lars Arge, and Jeffrey~Scott Vitter.
\newblock Bkd-tree: A dynamic scalable kd-tree.
\newblock {\em Lecture Notes in Computer Science}, 2750:46--65, 2003.

\bibitem{DBLP:conf/sigmod/Robinson81}
John~T. Robinson.
\newblock The k-d-b-tree: {A} search structure for large multidimensional
  dynamic indexes.
\newblock In {\em Proceedings of the 1981 {ACM} {SIGMOD} International
  Conference on Management of Data, Ann Arbor, Michigan, USA, April 29 - May 1,
  1981}, pages 10--18, 1981.

\bibitem{DBLP:books/wi/OkabeBSCK00}
Atsuyuki Okabe, Barry Boots, Kokichi Sugihara, Sung~Nok Chiu, and D.~G.
  Kendall.
\newblock {\em Spatial Tessellations: Concepts and Applications of Voronoi
  Diagrams, Second Edition}.
\newblock Wiley Series in Probability and Mathematical Statistics. Wiley, 2000.

\bibitem{DBLP:reference/db/ShahabiS18}
Cyrus Shahabi and Mehdi Sharifzadeh.
\newblock Voronoi diagrams.
\newblock In {\em Encyclopedia of Database Systems, Second Edition}. Springer,
  2018.

\bibitem{DBLP:reference/gis/ShahabiS17}
Cyrus Shahabi and Mehdi Sharifzadeh.
\newblock Voronoi diagrams for query processing.
\newblock In {\em Encyclopedia of {GIS.}}, pages 2446--2452. Springer, 2017.

\bibitem{Delaunay1934Sur}
B.~Delaunay.
\newblock Sur la sph\`ere vide. a la m\'emoire de georges vorono\"\i.
\newblock {\em Bulletin de I'Acad\'emie des Sciences de I'URSS. Classe des
  Sciences Math\'ematiques et Naturelles}, 6:793--800, 1934.

\end{thebibliography}
\vspace{12pt}

\end{document}